\documentclass[aps,prd,preprintnumbers,showpacs]{revtex4}
\usepackage[dvips]{graphicx}

\begin{document}

%
%

\eprint{Nisho-1-2022}
\title{Radiation Burst by Axion Star Collision with Star in the Andromeda Galaxy }
\author{Aiichi Iwazaki}
\affiliation{International Economics and Politics, Nishogakusha University,\\ 
6-16 3-bantyo Chiyoda Tokyo 102-8336, Japan }   
\date{Mar. 15, 2022}
\begin{abstract}
Axion is a promising candidate of dark matter in the universe.
A fraction of dark matter axion may forms axion star with radius $\sim 10^2$km.
We show that
the axion star emits radiation burst by the collision with K and M types main sequence star in the Andromeda Galaxy.
The emission arises in the atmosphere of the star, in which electrons
coherently oscillate due to oscillating electric field of the 
axion star. The electric field is produced under magnetic field $B$ of the star.
We estimate the flux density of the radiation 
$\sim 1.6\times 10^{-3}\mbox{Jy} (10^{-12}M_{\odot}/M_a)^2(10^{-5}\mbox{eV}/m_a)^3(B/10^2\mbox{G})^2\sqrt{3\times10^3\mbox{K}/T}$
and the rate of the collision per hour $\sim 0.06/\mbox{hour}\,(10^{-12}M_{\odot}/M_a)$ in the galaxy,
where $M_a$ ( $m_a$ ) denotes the mass of axion star ( axion ) and $T$ does temperature of the electrons.
We assume the 
number $10^{11}$ of the stars with $B\sim 10^{2}$G and radius $\sim 3.5\times10^{5}$km in the galaxy. 
We also assume that a half of the dark matter is composed of axion star. We show that
the emission of the radiation burst only arises in the atmosphere in which the plasma frequency $m_p\simeq m_a$.
The duration of the burst lasts for the period which it takes the axion star to pass the region with $m_p\simeq m_a$.
It would be longer than $1$ second.
  
\end{abstract}
\hspace*{0.3cm}

\hspace*{1cm}

\maketitle


Axion is the Goldstone boson of Peccei Quinn symmetry\cite{axion}, which naturally solves strong CP problem.
The axion is also a promising candidate of dark matter in the universe.
Therefore, its detection\cite{admx,carrack,haystac,abracadabra,organ,madmax,brass,cast,sumico,iwazaki01} 
is urgent issue for exploring beyond standard model. 
Such an axion is called as QCD axion, which was originally introduced to solve strong CP problem.
We only consider the QCD axion whose mass is 
constrained to the range\cite{Wil} of approximately $10^{-6}\mbox{eV} \sim10^{-3}$ eV. 
Especially, we assume that Peccei Quinn symmetry is spontaneously broken after inflation
in early universe. So, axion minicluster\cite{minicluster,minicluster1} may form and subsequently axion star\cite{axionstar,axionstar1,axionstar2,review} forms. 
Therefore, there is a possibility that a fraction of the dark matter axions form boson stars\cite{bosonstar1,bosonstar2,bosonstar3,bosonstar4}, 
i.e. axion stars.

\vspace{0.1cm}
In this paper we propose a new way of detection of the axion star. 
The detection is to observe radiation burst from the Andromeda Galaxy.
We consider radiations emitted by the axion star when it collides main sequence star
in the galaxy.
The axion star we consider is stable and a gravitationally loosely bound state of axions ( dilute axion star ). 
That is, binding energy of axion
is much less than the axion mass $m_a$. The mass of the axion star is supposed to be of the order of $10^{-12}M_{\odot}$.
The radius $R_a$ of such an axion star is approximately given\cite{radius,iwazaki}
such that $R_a\simeq1/Gm_a^2M_a$ with gravitational constant $G$ and mass $M_a$ of axion star.
Numerically, $R_a\simeq 260\mbox{km} (10^{-12}M_{\odot}/M_a)(10^{-5}eV/m_a )^2$.
Such axion stars may collide ordinary stars made of baryons.
Because the axion star is a coherent state of axions, the axion star generates classical oscillating electric
field\cite{iwazakiold} under the magnetic field of the main sequence star. The frequency of the oscillation is just given by the axion mass. 
Then, electrons in the atmosphere of the star are forced to oscillate by the electric field and coherently emit radiations\cite{coherentradiation}.
Because the coherent length $R_a\sim 100$km is large, 
a large number of electrons $N_e\propto n_eR_a^3\sim 10^{32}(n_e/10^{10}\rm cm^3)$ coherently emit radiations so that
their flux $\propto N_e^2$ can be extremely large.
The argument is very naive. We give more accurate estimation of the flux by solving Maxwell equations
and equation of motion of electron under both electric fields induced by the axions and of the emitted radiations. 
We confirm that the naive argument is essentially correct, although we have additional small factor in the flux coming
from small velocity of radiation in the electron plasma. 

\vspace{0.1cm}
Radiations from axion dark matter under stellar magnetic field 
have been previously discussed\cite{popov,coherentradiation,tka,hook,ben,hama,bu,pr,am}. 
Main emission mechanism
of radiations is non resonant conversion of axions into radiations, that is, 
incoherent conversion of an axion into a photon. On the other hand, resonant conversion of dark matter axion
have been also discussed\cite{hook,ben} in neutron star. 
But, the flux of such radiations are much 
smaller than the radiations discussed in this paper. 
The flux discussed previously is small 
because of non resonant conversion or small energy density of uniformly distributed axion dark matter.
Our point is 
the coherent emission by a large number of electrons under axion star: 
Energy density of axion star is $10^{24}$ times bigger than that of uniformly distributed axion dark matter. 
The mechanism is essentially the same as resonant conversion\cite{resonant,bu} of axion into radiations.
Photon in plasma acquires effective mass of plasma frequency $m_p$ and mixes with  
the axion in external magnetic field. Then, the resonant conversion of the axion into radiations occurs
only when $m_p\simeq m_a$.   
In this paper, by exactly solving Maxwell equations involving electric current of plasma, 
we show that a large amount of radiations arise only when $m_p\simeq m_a$.
This is similar to the resonant conversion of axion into radiations.
We make clear the relation between resonant conversion\cite{resonant,bu} of axion into radiations
and coherent emission of radiations by electrons\cite{coherentradiation}.
The resonant conversion and the coherent emission are intimately related each other.

\vspace{0.1cm}
Actual structure of magnetic field in the atmosphere of main sequence star is not well known.
In this paper we assume main sequence star whose surface magnetic field is $\sim 10^2$G and radius is $0.5R_{\odot}\simeq 0.35\times10^6$km.
We also assume that the magnetic field does not vary in time at least while the axion star emits large amount of radiations.
Such a stronger magnetic field than that of the sun is observed in K and M types main sequence star\cite{Mstar}.
Probably, they are main components in the Milky Way Galaxy and Andromeda Galaxy. Especially, we suppose that
the number of the stars colliding the axion star in the Andromeda Galaxy is of the order of $10^{11}$.
There is a perspective that there are such main sequence stars with $B\sim 10^2$G because 
white dwarfs are known to have strong magnetic field $\sim 10^6$G. 
Their magnetic fields are originated in the magnetic field of main sequence stars with $B\sim 10^2$G.

In general, radiation flux is larger as magnetic field is stronger.
Although radiation flux emitted in the axion star collision with stars of stronger magnetic field like
white dwarfs ( $B=10^6$G ) or neutron stars ( $B=10^{12}$G )\cite{coherentradiation,bu} is much larger,
the rate of the collision\cite{coherentradiation,bu} is much small to not be detectable.
Thus, we consider the axion star collision with the main sequence stars of the magnetic field $10^2$G
in the Andromeda Galaxy. Their number is much larger than those of white dwarfs and neutron stars.

\vspace{0.1cm}
When the axion star collides a main sequence star, at first it enter the atmosphere of the star.
The electron number density $n_e$ of the atmosphere as well as the strength of magnetic field gradually increase 
in the view of the axion star. The plasma frequency $m_p=\sqrt{e^2n_e/m_e}$ with electron mass $m_e$ also increases.
Accordingly, the radiation flux emitted by the electrons also gradually increases.
But the radiation flux is suddenly enhanced when the axion star passes the region with
the electron number density being nearly equal to a critical number density $n_e^{cr}$.
That is, the region with $m_p\simeq m_a$. The resonant conversion of axion into radiations occurs.
It should be noted that 
such an emission of a large amount of radiation stops 
because the axion star soon enters the region with $n_e>n_e^{cr}=m_em_a^2/e^2$, in other words $m_p>m_a$.
The radiations emitted in dense atmosphere with $m_p>m_a$
never arrive the earth. They are absorbed in the atmosphere. Thus, the maximum flux of the radiations we observe
is produced when the axion star passes the region with the critical electron number density 
$n_e^{cr}=m_em_a^2/e^2$ in the atmosphere. 
The duration of the radiation burst lasts for the period during which the axion star passes the region with 
$m_p\simeq m_a$.

\vspace{0.1cm}

Axion star we consider is a loosely bounded state of axions ( dilute axion star\cite{axionstar,iwazaki} ). 
There are no self interactions except for the mass term
because the field $a(t,\vec{x})$ is much small compared with the decay constant $f_a$.  
It is only bounded by 
gravitational force.  
( The axion field $a(t,\vec{x})$ describes the angle $\theta \equiv a(t,\vec{x})/f_a$ of complex scalar field $\Phi\propto \exp(i\theta)$. 
Thus, the angle $\theta \ll 1$ in the dilute axion star is very small. The overall scale of the field $a(t,\vec{x})$ is fixed by $f_a$. )

The field of the spherical axion star\cite{iwazaki,review} $a(t,\vec{x})=a(r)\cos(\omega t)$
is described by Schrodinger- Poisson equation,

\begin{equation}
-\frac{k^2}{2m_a}a(r)=-\frac{\vec{\partial}^2}{2m_a}a(r)+m_a\phi(\vec{x})a(r)
\end{equation}
with $\omega^2=m_a^2-k^2$ ( $k \ll m_a$ ) and $r=|\vec{x}|$,
where the gravitational potential $\phi(\vec{x})$ satisfies

\begin{equation}
\vec{\partial}^2\phi(\vec{x})=2\pi Gm_a^2a(r)^2
\end{equation}
and it behaves such as $\phi(\vec{x})\simeq -GM_a/r$ for $r\to \infty$.
The mass $M_a$ of the axion star is given such that 
$M_a=\int d^3x(\omega^2 a(r)^2+m_a^2a(r)^2)/4\simeq \int d^3x \,m_a^2a(r)^2/2$.
 
The Schrodinger- Poisson equation is approximately rewritten for $r \to \infty$ such that

\begin{equation}
-\frac{k^2}{2m_a}a(r)=-\frac{1}{2m_a}(\partial_r^2+\frac{2\partial_r}{r})a(r)-\frac{Gm_aM_a}{r}a(r)  
\end{equation}
Then, we find that
the spherical solution of the star is approximately described by $\tilde{a}(t,r)=f_a a_0\exp(-r/R_a)\cos(\omega t) $
with $k=1/R_a$,
where the radius $R_a$ of the axion star is

\begin{equation}
R_a=\frac{1}{Gm_a^2M_a}
\end{equation}
with $M_a=\pi m_a^2f_a^2a_0^2R_a^3/2$. 

We note that the radius $R_a$ is inversely proportional to the mass $M_a$. Such an axion star with small mass $M_a\sim 10^{-12}M_{\odot}$ 
is known to be stable. 
Numerically, 

\begin{equation}
\label{R}
R_a=260\mbox{km}\frac{10^{-12}M_{\odot}}{M_a}\Big(\frac{10^{-5}\mbox{eV}}{m_a}\Big)^2.
\end{equation}

( Although more detail analysis\cite{radius} shows larger radius than the formula $R_a=1/Gm_a^2M_a$, we use the formula in eq(\ref{R}).
The essence of our result does not change even if we use the larger radius. )   
\vspace{0.1cm}
When the axion star passes the atmosphere of star with magnetic field $B=10^2$G, the electric field is
induced on the axion star,

\begin{equation}
\label{e-field}
\vec{E}_a=-g_\gamma\alpha\frac{\tilde{a}(t,\vec{x})\vec{B}}{f_a\pi}\simeq -\frac{\alpha a_0 \vec{B}}{\pi}\cos(m_at)
\simeq 4.2 \times 10^{-28}\mbox{GeV}^2\cos(m_at)
\Big(\frac{M_a}{10^{-12}M_{\odot}}\Big)^2\Big(\frac{m_a}{10^{-5}\mbox{eV}}\Big)^3\frac{\vec{B}}{10^2\mbox{G}}
\end{equation}
with the fine structure constant $\alpha\simeq 1/137$ and we put $g_\gamma=1$, coupling strength between photon and axion
described below.
( The sign of the electric field is not important in our discussion. )

We explain how the electric field arises when the axion star is in the background magnetic field $B$.
It comes from the interaction such 
that the axion $a(t,\vec{x})$ couples with electric $\vec{E}$ and magnetic fields $\vec{B}$,

\begin{equation}
\label{L}
L_{aEB}= g_{\gamma}\alpha \frac{a(t,\vec{x})\vec{E}\cdot\vec{B}}{f_a\pi}\equiv g_{a\gamma\gamma} a(t,\vec{x})\vec{E}\cdot \vec{B}
\end{equation}
where the numerical constant $g_{\gamma}$ depends on axion models; typically it is of the order of one.
The standard notation $g_{a\gamma\gamma}$ is such that $g_{a\gamma\gamma}=g_{\gamma}\alpha/f_a\pi\simeq 0.14(m_a/\rm GeV^2)$
for DFSZ model\cite{dfsz,dfsz1} and $g_{a\gamma\gamma}\simeq -0.39(m_a/\rm GeV^2)$ for KSVZ model\cite{ksvz,ksvz1}.
In other words, $g_{\gamma}\simeq 0.37$ for DFSZ and $g_{\gamma}\simeq -0.96$ for KSVZ.
The axion decay constant $f_a$ is related with the axion mass $m_a$ in the QCD axion; $m_af_a\simeq 6\times 10^{-6}\rm eV\times 10^{12}$GeV.
We only consider the QCD axion in this paper.

\vspace{0.1cm}
The interaction in eq(\ref{L}) between axion and electromagnetic field slightly modifies Maxwell equations in vacuum

\begin{eqnarray}
\label{modified}
\vec{\partial}\cdot(\vec{E}+g_{a\gamma\gamma}a(t,\vec{x})\vec{B})&=0&, \quad 
\vec{\partial}\times \Big(\vec{B}-g_{a\gamma\gamma}a(t,\vec{x})\vec{E}\Big)-
\partial_t\Big(\vec{E}+g_{a\gamma\gamma}a(t,\vec{x})\vec{B}\Big)=0,  \nonumber  \\
\vec{\partial}\cdot\vec{B}&=0&, \quad \vec{\partial}\times \vec{E}+\partial_t \vec{B}=0.
\end{eqnarray}
From the equations, we approximately obtain the electric field 
$\vec{E}_a=-g_{a\gamma\gamma}\tilde{a}(t,\vec{x})\vec{B}\simeq -\alpha a_0 \vec{B}\cos(m_at)/\pi$ in eq(\ref{e-field})
generated by the axion star $\tilde{a}(t,\vec{x})$  
under background magnetic field $\vec{B}$.
It is of the order of $g_{a\gamma\gamma}\tilde{a}(t,\vec{x})$ ( $ \ll 1$. )

\vspace{0.2cm}
First of all, we give a naive argument how large radiations arise when the axion star collides electron gas.
In the argument we consider only electric field $\vec{E}_a$, which makes electrons oscillate.
Later, we discuss the radiations by taking account of not only the electric field $\vec{E}_a$ but also electric field $\vec{E}_{rd}$ of radiations
emitted by electrons themselves.
We find that the naive argument 
is essentially correct, although there is an additional much small factor suppressing radiation flux
in the electron plasma when $m_p\simeq m_a$.

\vspace{0.1cm}
The electric field $E_a$ makes an electron harmonically oscillate because it oscillates with frequency $m_a$.
Because there are many electrons in the atmosphere of main sequence star,
the oscillating electric field of the axion star induces coherent oscillation
of the electrons over the axion star. The coherent length is approximately given by $R_a$.
They emit coherent dipole radiations, i.e. coherent electromagnetic radiations 
with the frequency $m_a/2\pi\simeq 2.4\mbox{GHz}(m_a/10^{-5}\mbox{eV})$. 
Thus, the power of the radiation  $\dot{\omega}N_e^2$
is extremely enhanced because the number $N_e=4\pi n_eR_a^3/3$ of the electrons is large,
where $\dot{\omega}=2\alpha(eE_a)^2/3m_e^2$ denotes the radiation power emitted by a single electron with its mass $m_e$.
Numerically, the power of a single electron averaged in time is

\begin{equation}
\dot{\omega}=\overline{2\alpha(eE_a)^2/3m_e^2}=
3.8\times 10^{-31}\mbox{erg/s}\Big(\frac{M_a}{10^{-12}M_{\odot}}\Big)^4\Big(\frac{m_a}{10^{-5}\mbox{eV}}\Big)^6
\Big(\frac{B}{10^2\mbox{G}}\Big)^2
\end{equation} 
with time average $\overline{Q}$ of $Q$.

The radiations are mainly emitted into the direction perpendicular to the magnetic field $\vec{B}$ because electrons harmonically 
oscillate in the direction $\vec{E}_a$ parallel to $\vec{B}$. It implies that they are hardly affected by the magnetic field. 
The radiations can propagate to the outside of the atmosphere without absorption, in which the plasma frequency is lower than
the frequency $m_a$. 
 
\vspace{0.1cm}
When the axion star enters the atmosphere of a main sequence star in the collision, 
the flux of the radiations gradually increases as the electron number density increases.
But the radiations are
suddenly enhanced, when it passes the region with $m_p\simeq m_a$.
After the enhancement of the emission, the radiations stop because
it goes into the deep region of 
the atmosphere in which the plasma frequency $m_p=\sqrt{e^2n_e/m_e}$ is 
larger than the frequency $m_a$.
The critical number density is given by $n_e^{cl}\simeq 7\times10^{10}/\mbox{cm}^3(m_a/10^{-5}\mbox{eV})^2$.
The radiations are absorbed in such dense electron gas with $n_e>n_e^{cl}$.

\vspace{0.2cm}
Thus, the maximum power of the coherent radiations is given by

\begin{equation}
\label{N^2}
\dot{\omega}N_e^2\simeq 1.0\times 10^{30}\mbox{W}\Big(\frac{10^{-12}M_{\odot}}{M_a}\Big)^2
\Big(\frac{10^{-5}\mbox{eV}}{m_a}\Big)^2\Big(\frac{B}{10^2\mbox{G}}\Big)^2
\end{equation}
with $1\mbox{W}=10^7\mbox{erg/s}$,
where we assume that
the number density of electrons emitting the radiations is given by
$n_e^{cl}=7\times 10^{10}(m_a/10^{-5}\mbox{eV})^2/\mbox{cm}^3$. 
We note that $\dot{\omega}\propto M_a^4m_a^6B^2$ and
$N_e^2\propto n_e^2R_a^6\propto M_a^{-6}m_a^{-8}$.
The result is obtained with the naive argument which only takes into account the electric field $\vec{E}_a$ 
induced by the axion star. When we take into account electric field $\vec{E}_{rd}$ of the radiation emitted by electrons,
we have additional much small factor in the formula $\dot{\omega}N^2$.
Furthermore, we find that the enhanced radiation only arises in the region with $m_p\simeq m_a$.

\vspace{0.1cm}
Now, we would like to more accurately discuss the radiations by taking account of
back reactions of the emitted radiations on electrons.
An electron with velocity $\vec{v}$ is made to oscillate by electric fields according to the equation of motion,

\begin{equation}
\label{eq}
m_e\dot{\vec{v}}=e(\vec{E}_a+\vec{E}_{rd})
\end{equation}
where $\vec{E}_{rd}$ denotes the electric field of the radiations emitted by the electrons themselves.
We notice that electron receives the effects of two types of electric fields. One is the electric field $\vec{E}_a$ induced 
on the axion star under the magnetic field $\vec{B}$, while the another one is $\vec{E}_{rd}$.
In the naive argument above, we have not taken into account the effect of the electric field $\vec{E}_{rd}$.

Using Maxwell equations, $\vec{\partial}\times \vec{B}_{rd}=\vec{J}+\partial_t\vec{E}_{rd}$ 
and $\vec{\partial}\times \vec{E}_{rd}=-\partial_t\vec{B}_{rd}$ with electric current $\vec{J}=en_e\vec{v}$,
we find that the electric field $\vec{E}_{rd}$ satisfies the equation 

\begin{equation}
\label{eqE}
(\partial_t^2-\vec{\partial}^2+m_p^2)\vec{E}_{rd}=-m_p^2\vec{E}_a
\end{equation}
with the plasma frequency $m_p$; $m_p^2=e^2n_e/m_e$.
It can be also derived by using the equations(\ref{modified}) with
the current $\vec{J}$.
The equation in (\ref{eqE}) 
describes the emission of radiation produced by the electric current $\vec{J}$.

\vspace{0.1cm}
We note $\vec{E}_a\propto \vec{B}$. Thus,
supposing that $\vec{E}_a=(0,0,E_a)$ and $E_a=E_a^0 \exp(-r/R_a)\exp(-im_a t)$ with $E_a^0\equiv \alpha a_0 B/\pi$,
we find the solution $\vec{E}_{rd}=(0,0,E_{rd})$,

\begin{equation}
E_{rd}=\exp(-im_at)\Big(\frac{-p_2\exp(ikr)}{r}+\frac{p_1r+p_2}{r}\exp(-\frac{r}{R_a})\Big),
\end{equation}
with $k=\sqrt{m_a^2-m_p^2}$ ( we assume $m_a \ge m_p$. ), $p_1=E_a^0 m_p^2R_a^2/(k^2R_a^2+1)$ and $p_2=2E_a^0m_p^2R_a^3/(k^2R_a^2+1)^2$,
where $k$ denotes the momentum of radiation in the electron gas with plasma frequency $m_p$. 
The momentum $k$ is not equal to the frequency $m_a$ of radiation in the plasma.  
The coefficient $p_2$ of the term $\propto \exp(-im_at+ikr)$ is determined by the regularity at $r=0$ of the electric field $E_{rd}$.
The term represents the radiation propagating	 far from the axon star.

The magnetic field $\vec{B}_{rd}$ is also obtained using the formula, 
$\vec{\partial}\times \vec{E}_{rd}=-\partial_t\vec{B}_{rd}=im_a \vec{B}_{rd}$,

\begin{eqnarray}
\vec{B}_{rd}&=&\frac{1}{im_a}(\partial_x E_{rd},-\partial_y E_{rd},0)=\frac{1}{im_a}(\frac{y}{r},-\frac{x}{r},0)\exp(-im_at)\epsilon \\
\epsilon&\equiv &p_2(\frac{-ik}{r}+\frac{1}{r^2})\exp(ikr)-(\frac{p_a}{r^2}+\frac{p_1r+p_2}{R_ar})\exp(-\frac{r}{R_a})
\end{eqnarray}

\vspace{0.1cm}
The field strengths at the large distance $r \gg R_a$ behave as 

\begin{equation}
\vec{E}_{rd}\to \exp(-im_at+ikr)\frac{-p_2}{r}\big(0,0,1\big), \quad \vec{B}_{rd}\to \exp(-im_at+ikr)\frac{-kp_2}{m_ar}\big(\frac{y}{r},\frac{x}{r},0\big) 
\quad \mbox{for} \quad r \gg R_a
\end{equation}
Because $p_2\simeq E_a^0 m_p^2R_a^3\propto n_eR_a^3 \gg 1$ for $kR_a<1$, 
the radiation is coherently emitted by electrons with their number $N_e\sim n_eR_a^3$. The field strength is very large.
This is the situation mentioned above in the naive argument where $N_e$ electrons coherently oscillate 
by the electric field $\vec{E}_a$.
On the other hand, $p_2\sim E_a^0m_p^2k^{-4}R_a^{-1}\sim n_eR_a^3/(kR_a)^4 \ll n_eR_a^3$ for $kR_a \gg 1$. 
Thus, the radiation is suppressed.
The momentum range $k<1/R_a$ ( $k/m_a \sim 10^{-7}$ ) for the enhancement is extremely small. 
It implies that the emission of a large amount of the radiation occurs
only in such a narrow range. This is similar to resonant conversion of axion into radiations.

In other words, when the coherent length $1/k$ of the radiation $\vec{E}_{rd}\propto \exp(ikr)$ is larger than or similar equal to
the coherent length $R_a\sim 100$km of the electric field $\vec{E}_a\propto \exp(-r/R_a)$ i.e. $kR_a \le1$,
the coherent oscillation by $\vec{E}_a$ is not disturbed by the radiation $\vec{E}_{rd}$.
Thus, the coherent radiations arise.
This corresponds to the resonant conversion of axion into radiations. 
On the other hand, when 
$1/k$  is much smaller than $R_a$, such coherent radiations do not arise. The radiation field $\vec{E}_{rd}$ emitted by electrons
disturbs the coherent oscillation by $\vec{E}_a$. Thus,
the radiations emitted by the electrons
are very weak.

\vspace{0.1cm}
At the first glance, our emission mechanism of the radio burst is different from the emission due to axion resonant conversion.
Our mechanism requires coherent oscillation of electrons, while the resonant conversion does the mixing of photon and axion
with identical mass. When plasma frequency is equal to the axion mass, 
both wave lengths of the photon $1/k_{rd}$ and axion $1/k_a$ are identical, i.e. $k_{rd}=k_a$.
Thus, electric field induced by the axion makes electron coherently oscillate over the wave length $1/k_a$ and the coherent
oscillation is never disturbed by the electric field of the photon emitted by the electrons themselves.  
Thus, physically, the resonant conversion of axion into radiations takes place owing to
the coherent emission by electrons.

\vspace{0.1cm}
As we have obtained electromagnetic fields propagating at large distance, we can find the flux of the radiations,

\begin{equation}
\label{F}
F=\frac{1}{2}\int d\vec{S} \cdot(\vec{E}_{rd}\times \vec{B^{\dagger}}_{rd})=\frac{4\pi k}{3m_a}p_2^2
=\frac{4\pi k}{3m_a}\Big( \frac{2E_a^0m_p^2R_a^3}{(k^2R_a^2+1)^2}\Big)^2 \simeq \frac{\pi k}{3m_a}(E_a^0 m_p^2R_a^3)^2
=\frac{9}{4}\frac{k}{m_a}\dot{\omega}N_e^2
\end{equation}
with $kR_a\simeq 1$ ( $m_p\simeq m_a$ ),
where we consider the region with plasma frequency $m_p$ being nearly equal to $m_a$
in which the flux is enhanced due to the coherent emission by $N_e=4\pi n_eR_a^3/3$ electrons. 

\vspace{0.1cm}
We should notice that the flux in eq(\ref{F}) is different with the naive one in eq(\ref{N^2}) by the factor
$k/m_a\simeq k/m_p$. The factor $k/m_p=1/m_pR_a \sim 10^{-7}(260\mbox{km}/R_a)(10^{-5}\mbox{eV}/m_a)$ describes the velocity of the radiations
inside the plasma. Thus, the flux is much smaller than the naive one because the velocity is much small;
the naive estimation assumes the light velocity of the radiations in the vacuum.

\vspace{0.1cm}
We now proceed to examine the radiation flux from the Andromeda Galaxy in which 
the axion star collides main sequence star with magnetic field $\sim 10^2$G.
Because the distance to the Andromeda Galaxy is $\sim 2.5\times 10^6\rm ly$,
the observed flux density $S_{\nu}$ of the radiation is estimated in the following,

\begin{equation}
S_{\nu}=\frac{F}{4\pi(2.5\times 10^6\mbox{ly})^2}\frac{1}{\delta \nu}\simeq 1.6\times 10^{-3}\mbox{Jy}\Big(\frac{g_{\gamma}}{1.0}\Big)^2
\Big(\frac{10^{-12}M_{\odot}}{M_a}\Big)\Big(\frac{10^{-5}\mbox{eV}}{m_a}\Big)^2
\Big(\frac{B}{10^2\mbox{G}}\Big)^2\sqrt{\frac{3\times 10^3\mbox{K}}{T}}
\end{equation}
with $1$Jy$=10^{-26}\mbox{W}/(\rm m^2 Hz)$, and $g_{\gamma}=0.37$ ( $g_{\gamma}=-0.96$ ) for DFSZ ( KSVZ ),
where the bandwidth $\delta \nu$ of the line spectrum is caused by the temperature $T$ of electron gas:
$\delta \nu=(m_a/2\pi)(\sqrt{T/m_e})\simeq 1.7\mbox{MHz}(m_a/10^{-5}\mbox{eV})\sqrt{T/(3\times10^3\rm K)}$. 
The line spectrum is thermally broadened.
We tentatively use the value $T=3\times 10^3$K of the temperature,
although real ones would be different in each collision. ( We consider red dwarf with low surface temperature such as $T=3\times10^3$K. )
The emission of the burst lasts for the time in which it takes for the axion star to pass such a region with $m_p\simeq m_a$. 
It would be more than $1$ second because the velocity of the axion star is of the order of $3\times 10^2$km/s.

\vspace{0.1cm}
When the axion star collides a main sequence star, at first it enter the atmosphere of the star in which
the plasma frequency is much smaller than $m_a$ and the radiation from electrons is very weak because $k\simeq m_a$
or $kR_a \gg 1$.
Then, it proceeds to deep inside the atmosphere and reaches the critical region with $m_p \simeq m_a$,
in which coherent radiations with the large flux in eq(\ref{F}) are emitted.
In other words, 
the momentum $k$ ( velocity ) of the radiations gradually decreases as the plasma frequency increases.
Finally, enhanced coherent radiations are emitted in the critical region with  $m_p \simeq m_a$,
where the momentum ( velocity $=k/m_p$ ) is very small, $k \sim R_a^{-1}$.

%

\vspace{0.1cm}
We now estimate the event rate of the axion star collisions with main sequence stars in the Andromeda Galaxy. 
Especially we consider K and M types star because they would make up more than a half of stars in the galaxy and
they have strong surface magnetic field\cite{Mstar} of the order of $10^3$G.
We suppose that 
the number of $N_s=10^{11}$ such stars collide the axion star  
and that the fraction of the axion star in the dark matter is $1/2$. That is, the number density $n_a$ of the axion star is given by
$n_a=0.5\times \rho_d/M_a$ where we denote the dark matter density as $\rho_d$. It is assumed that $\rho_d\simeq 0.3\mbox{GeV}/\mbox{cm}^3$.
Thus, we estimate the event rate of the collisions per hour in the Andromeda galaxy is

\begin{equation}
\label{rate}
\mbox{event rate of collision per hour} =
n_a\times N_s\times Dv \simeq 0.06/\mbox{hour} \Big(\frac{10^{-12}M_{\odot}}{M_a}\Big)
\Big(\frac{\rho_a}{0.5\rho_d}\Big)\Big(\frac{N_s}{10^{11}}\Big)
\end{equation}
with relative velocity $v=10^{-3}$ between axion star and the main sequence star,
where the collision cross section $D$ is given by

\begin{equation}
D=\pi(R_s+R_a)^2\Big(1+\frac{2GM_s}{v^2(R_s+R_a)}\Big)\simeq \pi R_s^2\Big(1+\frac{2GM_s}{v^2R_s}\Big)\simeq 5.3\pi R_s^2,
\end{equation}  
supposing radius $R_s=0.5R_{\odot}\simeq 3.5\times 10^5$km and mass $M_s=0.5M_{\odot}$ of the colliding star.
Therefore, such a collision between the axion star and K, M type main sequence star in the Andromeda Galaxy occurs so frequently 
that we can observe the collision
using existing radio telescopes.

\vspace{0.1cm}

We have several comments on the above discussions.

First,
we have assumed the mass of the axion star $M_a\sim10^{-12}M_{\odot}$.
Such an axion star is stable\cite{radius}. 
( The critical mass beyond which axion stars are unstable is approximately given by $10^{-11}M_a$
for $m_a=10^{-5}$eV. ) 
The mass is consistent with the estimation\cite{minicluster,minicluster1} of the mass of axion minicluster.
The minicluster is discussed\cite{lev,egg} to become axion star by axion condensation.
It is reasonable to use the mass. 
Additionally, it is proposed\cite{coherentradiation,iwazaki1,bu,iwazaki2} that fast radio bursts arise by axion star collisions with neutron star.
The event rate of the collisions is consistent with the burst event when $M_a\sim 10^{-12}M_{\odot}$.
These are the reason for our choice of the mass $M_a\sim10^{-12}M_{\odot}$.

\vspace{0.1cm}
Secondly,
The above estimation involves some ambiguities. 
The most ambiguous one is the assumption of number density $n_a=0.5\times \rho_d/M_a$
of axion star. It might be a fraction of the dark matter such as $n_a\sim 0.1\times \rho_d/M_a$.
Furthermore, 
we have used the parameter $B=10^2$G. But,
in general, K and M types main sequence stars would possess 
stronger magnetic field than $10^2$G. Then,
the flux density $S_{\nu}$ is much larger than the estimation $\sim 10^{-3}\mbox{Jy}(B/10^2\rm G)^2$.
For instance, $S_{\nu}\sim 10^{-1}$Jy $(B/10^3\rm G)^2$.
The signal of the collision is much clearer. 
Additionally, in general,
we have no detailed observation of magnetic field configuration and the distribution of electron plasma in star.
Thus, we cannot definitely anticipate the duration of the radio burst. It is given by the time which it takes for axion star
to pass the region with $m_p\simeq m_a$ and $B\simeq 10^2$G.

\vspace{0.1cm}
Thirdly, although we have examined radio bursts arising the Andromeda Galaxy, we may consider the bursts arising in our Milky Way Galaxy.
Then, the radiation flux of the bursts are $4$ orders of magnitude larger than that in the Andromeda Galaxy
because the distance to the burst in our galaxy is of the order of $10^2$ times smaller than that to burst in the Andromeda Galaxy.
Thus, we expect large flux density $S_{\nu}\sim 10\rm Jy(B/10^2\rm G)^2$, although the detection efficiency is much small.
We cannot observe the whole of our galaxy in one time.

\vspace{0.1cm}
Finally,
we would like to make a comment on the deformation of the axion star by tidal force caused by the main sequence star.
The deformation begins to occur when the axion star approaches the star within the distance\cite{iwazaki3} 
$R_c=R_a(0.5M_{\odot}/M_a)^{1/3}\simeq 1.9\times 10^6$km. The distance is roughly equal to the radius $R_s=3.5\times10^5$km of the star.
Thus, the use of the spherical form $\tilde{a}(t,r)$ of the axion star is good approximation. The deform from the spherical form
is not so large for the deformation to make an effect on the above estimation.

\vspace{0.2cm}
To summarize, our emission mechanism ( radiation burst from axion star collision with magnetic 
main sequence star ) is a promising way for the observation of the axion star. 
Indeed, 
the axion star collision with K and M types main sequence stars in the Andromeda Galaxy frequently occurs $\sim 1/\rm day$.
The flux density  $\sim 10^{-1}$ Jy$(B/10^3\rm G)^2$ of the radiation burst would be large enough for the detection
if the duration is longer than $1$ second. 
The burst shows line spectrum with very small width $\delta\nu\sim 1$MHz.
Therefore, the burst may be detectable with existing radio telescopes. The observation of the burst
reveals the existence of the axion and determines the value of the axion mass.

\vspace{0.2cm}
The author
expresses thanks to S. Kisaka for useful information and
members of theory group in KEK for their hospitality.
This work is supported in part by Grant-in-Aid for Scientific Research ( KAKENHI ), No.19K03832.



\end{document}